\documentclass{elsart}

\usepackage{amssymb}

\begin{document}

\begin{frontmatter}



\title{On the potential catastrophic risk from 
metastable quantum-black holes produced at particle colliders}

\author[label2]{R. Plaga}

\ead{rainer.plaga@gmx.de}

\address[label2]
{Franzstr. 40, D-53111 Bonn, Germany}

\begin{abstract}
The question of whether collider produced
of subnuclear black holes
might constitute a catastrophic risk is explored in a model of Casadio \& Harms (2002)
that treats them as quantum-mechanical objects.
A plausible
scenario in which these black holes accrete ambient matter
at the Eddington limit shortly after their
production, thereby emitting Hawking radiation 
that would be harmful to Earth and/or CERN
and its surroundings, is described. 
Such black holes are shown to 
remain undetectable in existing astrophysical observations and thus evade
a recent exclusion of risks from subnuclear black holes by Giddings \& Mangano (2008) and
and a similar one by Koch et al. (2009).
I further question that these risk analyses
are complete for the reason that they exclude 
plausible black-hole parameter
ranges from safety consideration without giving
any reason. Some feasible operational measures at colliders
are proposed
that would allow the lowering of any remaining risk probability.
\\
Giddings \& Mangano drew 
different general 
conclusions only because they made
different initial assumptions about the properties
of microscopic black holes, not because
any of their technical conclusions are incorrect.
A critical comment by 
Giddings \& Mangano (2008) on the present paper and 
a preprint by Casadio et al.(2009) - that
presents a treatment of the present issue with methods
and assumptions similar to mine -
are addressed in appendices.
\end{abstract}

\end{frontmatter}
\section{Introduction}
\subsection{General outline of the problem}
Theories with ``extra spatial dimensions''\cite{landsberg}, are one 
of the most popular extensions of the
standard model of particle physics and a central plank of string theory\cite{forste}. 
If space had more than three dimensions the ``Planck energy scale'' - usually
thought to lie at extremely high energies - could 
be reached already at energies projected for new particle accelerators\cite{ADD,randall}.
Gravity becomes strong
near the Planck energy. As a consequence
subnuclear ``micro''
black holes (mBHs) could be copiously produced at future 
high-energy particle colliders\cite{dimopoulos,landsberg}, such as the
``Large Hadron Collider'' (LHC) at CERN.
At a predicted rate of up to about one BH per second at the nominal
LHC luminosity\cite{dimopoulos}, the LHC would be a ``black-hole factory''\cite{giddings,barrau}.
The phenomenology of mBHs at colliders 
has been studied in great detail, see e.g. Cavaglia et al.\cite{cav}.
\\
The possibility that a collider-produced
black hole (BH) - or another exotic object -
might catastrophically grow by gravitationally pulling in ambient matter and thus
eventually injure or kill humans deserves careful attention\cite{wil,kent,calogero,rees}.
A recent scientific comparative
study of global risks\cite{leggett} has put a risk very similar to the one
considered here (from collider-produced ``strangelets'') at the top
``response priority'' of all current ``untreated risks'' 
(such as, for example, super-volcano
eruptions and asteroid impacts).
Clearly this potential risk exists only if speculative theories are correct.
But these theories were constructed to explore real possibilities. 
The probability that
they are correct is not negligible.
\subsection{Definition of risk scenarios}
\label{risk_scen}
\subsubsection{The scenarios discussed by Giddings \& Mangano}
Recently this risk has been studied in great detail
in important papers by Giddings \& Mangano (G \& M)\cite{gm}
\footnote{
The conclusions of a report\cite{lsag} by the ``lsag group'' at CERN on the safety 
of microscopic black holes are entirely based on results from this paper.}
and Koch et al.\cite{koch}.
G \& M consider two frameworks for the description
of mBHs. In a {\it ``first scenario''} (case ``D0'' in 
Koch et al.\cite{koch})
collider-produced mBHs are treated in a popular, standard manner
with a semiclassical 
thermodynamical description (i.e. assuming a
canonical ensemble). The mBH is described as
a heat bath and any back reaction of the emitted particles
on the mBH is neglected.
mBHs are then expected to 
decay, via the emission of
``Hawking radiation'', on extremely small timescales after their production,
thus they cannot grow and pose no danger. 
\\
In a {\it second scenario} (case ``D1-B'' in Koch et al.\cite{koch})
G \& M assume that mBHs ``do not undergo Hawking decay'' 
in a purely ad hoc manner, in order to ``conduct an independent
check of their benign nature''. They further plausibly assumed that
mBHs shed any electrical charge they acquired due to accretion 
very rapidly via the Schwinger mechanism.
They rightly point out that this {\it scenario 2}, while
not being completely unphysical\footnote{G \& M quote Unruh \& Sch\"utzhold\cite{unruh,unruht}
who constructed a speculative model potentially without Hawking radiation.}, is not
preferred ``on very general grounds''. G \& M
study the behaviour of mBHs after their production at the LHC in this scenario and
find that for certain possible choices of parameters a collider produced mBH
might accrete Earth on time scales, quote, ``that are too short to provide comfortable
constraints''. The existence of mBHs within this ``dangerous'' parameter range is then excluded by
making the case that cosmic-ray produced mBHs 
would accrete certain observed white dwarfs with small magnetic fields
on smaller time scales than their age\footnote{G \& M argue that
neutron stars might survive on time scales
comparable to their observed age, so that their
existence does not compellingly rule out the existence
of ``dangerous'' black holes.}. G \& M find that the fact that our Earth, the sun and
other stars were not destroyed by ``dangerous'' cosmic-ray produced black holes does not
exclude their existence because they 
would not be stopped within them after production.
Koch et al.\cite{koch} use similar arguments and reach the same conclusions.
\subsubsection{A new scenario not discussed previously by G \& M or Koch et al.:{\it Scenario 3}}
\label{scen3}
It is the aim of the present paper to explore a {\it third
scenario}, in addition to the two presented by G \& M and the
ones by Koch et al.\cite{koch}\footnote{Koch et al.
quote the present paper in connection with their
case ``D1-B'' (which corresponds to {\it scenario 2}). 
If they thought that the present paper
is about this case, they erred. Moreover they present three additional
scenarios in which mBHs are electrically charged and turn out to be ``harmless''.}.
Its basic difference to {\it scenario 1} is a completely different ``microcanonical''
treatment of mBH thermodynamics (i.e. one
in which the total energy remains fixed)\cite{mc,casadioh01,rizzo2,ging}, leading to a strongly
reduced, but not completely switched-off intensity of Hawking radiation for
black holes with small masses.
This treatment is thought
to be more fundamental than the one from G \& M's {\it scenario 1} 
using the canonical ensemble.
The mBHs are typically described as extended stringy objects, like
e.g. ``p-branes''. They are then a new type of elementary particle
a ``quantum black hole''\cite{hl93,keski,wilczek}. 
It is still assumed that mBHs neutralize quasi instantenously
via the Schwinger mechanism.
\\
In a sense such a framework is more plausible than 
both frameworks studied by G \& M because
it can avoid a violation of
unitary evolution and energy conservation\cite{chl,callan}, serious problems that
are well known to beset the {\it first scenario} used by G \& M\cite{parikh,gm}. 
Moreover it is not ``ad hoc'' such as {\it scenario 2}, but based on models
published in the peer-reviewed literature.
G \& M endorse
a quantum mechanical treatment of mBHs at the end of their section 2.1, but they
do not develop this possibility further in their report. 
\\
{\it Scenario 3} adopts 
Casadio \& Harms' (C \& H)\cite{casadio} model for mBHs.
The famous ``Randall-Sundrum 2 (RS 2) model''\cite{randall} - presented
in one of the most frequently quoted papers in
the recent history of high energy physics - is chosen as the description
of a 4th spatial dimension\footnote{Thus I follow the
treatment in section III.B of C \& H\cite{casadio}.}.
In trying to understand if mBHs could be dangerous in {\it scenario 3} I will repeatedly resort
to a use of G \& M's excellent theoretical tools. I try to 
assume reasonably mild worst case parameter choices, similar to the strategy
of G \& M\footnote{G\& M wrote: ``...at each point where we encountered an
uncertainty, we have replaced it by a conservative or ``worst case'' assumption''.}.
However, I strove to introduce no ``ad hoc'' or fine tuned assumption, that would 
be deemed highly implausible to experts.
\\
Section \ref{ch} reviews the Hawking luminosity
of mBHs within 
{\it scenario 3}. Section \ref{hdes} explains why - for
certain parameter ranges - {\it scenario 3} predicts a disaster that is not
ruled out by astrophyiscal considerations. It also explores if
such disasters could be of limited scale. Section \ref{danger} points out
a gap in 
the astrophysical ``safety arguments'' of G \& M that is independent of the consideration
in section \ref{hdes}.
Finally section \ref{concl} concludes. Appendix \ref{gm_crit}
answers a critical comment of G \& M on the present paper and appendix \ref{cas_paper} comments
recent work of Casadio et al. on the present issue.

\section{Properties of RS 2 quantum microscopic black holes in the Casadio \& Harms model}
\label{ch}
\subsection{Introduction}
It seems likely that
quantum black holes are in principle unstable, i.e. they 
eventually evaporate by Hawking
radiation because no conserved
quantum number forbids them to do so\cite{gm}. However
within the microcanonical treatment
of black holes developed by R.Casadio, B.Harms and Y.Leblanc \cite{chl}
(assumed in my {\it scenario 3}),
if their mass is smaller than a certain mass scale ``M$_N$''
of their theory,
they live much longer than expected within the standard
thermodynamic treatment that was employed for G \& M's {\it scenario 1}.
Therefore - in contradistinction to G \& M's {\it scenario 1} - we can neither simply
assume that the mBHs evaporate before they can do harm, nor 
that there is no potentially dangerous Hawking radiation (as in G \& M's {\it scenario 2}).
Rather we will
need to study their fate after production taking into account 
both accretion and possible effects from Hawking radiation in section \ref{hdes}.
As a preparation I review the intensity of Hawking radiation of quantum mBHs 
in {\it scenario 3} in this section.

\subsection{Stability of microscopic black holes for various possible input parameters}
If the additional curved spatial dimension of the RS 2 model exists,
C \& H predict a Hawking luminosity of the mBH of\cite{casadio}: 
\begin{equation}
P_5 = {{M_{\rm BH} \hbar c^6} \over {15360 \pi G^2 M_N^3}} 
\label{lumi5}
\end{equation}
Casadio \& Harms\cite{casadio} apply this formula to black holes with a mass M$_{\rm BH}$ 
smaller than the parameter ``M$_N$'' within their theory.
I follow them and all calculations in my section \ref{hdes} below assume this
relation.
For black-hole masses exceeding M$_N$  C \& H  assumed the
classical canonical 4-dimensional expression for the Hawking luminosity: 
\begin{equation}
P_4 = {{ \hbar c^6} \over {15360 \pi G^2 M_{\rm BH}^2}}.
\label{lumi4}
\end{equation}
The luminosity of eq.(\ref{lumi5}) is normalized to the
classical expression at the mass M$_N$ (eq.(\ref{lumi5}) is
already normalized in this way).
The critical difference of this treatment to the usual, canonical one is exactly that
the new microcanocial ``quantum'' expression eq.(\ref{lumi5}) has to be employed.
\\
For given curvature scale ``L'' (a length scale associated
with the warping in the RS 2 model) C \& H assumed
that M$_N$ is equal to a
black hole mass at which Schwarzschild radius
of a 5-dimensional mBH
reaches L. This gives:
\begin{equation}
M_N = {{3 \pi L^2 c^2 M_5^3}\over{8 \hbar^2}}  
\label{norm1}
\end{equation}
Here M$_5$ is the ``new'' Planck scale
(set to 1 TeV in all numerical estimates below).
Because P$_5$ = P$_4 {M_{\rm BH}^3 \over M{_N}^3 }$,
the Hawking luminosity of black-holes with initial masses (typically 10$^{-23}$ kg)
much below M$_N$ 
(possibly $\gg$ kg, see section \ref{casa2}) 
is strongly suppressed with respect to the classical 
value P$_4$\footnote{G \& M do not deny that the Hawking luminosity
of 5-dimensional black holes is suppressed with 
respect eq.(\ref{lumi4})\cite{gm}, but the suppression is weaker
in {\it scenario 1} than it is in {\it scenario 3}.}.
However, with growing mass (e.g. by accretion) the suppression
of the Hawking radiation is lifted.
\\
The geometry of mBHs with Schwarzschild radii
between L and $\approx$ 6 $\times$ L\footnote{This range was
derived from eq.(3.26) of G $\&$ M for the
scales of L of interest in this manuscript: 10$^{-9}$ m $<$ L $<$ 10$^{-4}$ m.
If L was smaller, mBHs in the third scenario would pose no
catastrophic risk, because they would decay faster than they would grow.}
is not known, and it remains presently unclear if 
eq.(\ref{lumi4}) can be applied in this ``transitional region'' as assumed
above.
Only for black holes with masses above ``M$_C$'', the mass
of a mBH with a Schwarzschild radius of 6L, above
which a 4-dimensional description of the mBH is a good approximation,
does this appear to be certain. M$_C$ is given as:
\begin{equation}
M_C \approx {{3 L c^2}\over{G}}
\end{equation}
Thus one might equally well normalize the luminosity equally M$_C$ setting:
\begin{equation}
M_N = M_C
\label{norm2}
\end{equation}
The decision between normalisation in eq.(\ref{norm1}) and eq.(\ref{norm2})
comes down to the question of whether the luminosity of a mBH is
described by the 5-dimensional (eq.(\ref{lumi5})) or 4-dimensional
(eq.(\ref{lumi4})) expression in the transitional region between
L and $\approx$ 6 L. All one can presently say
with reasonable certainty is that the correct normalisation lies 
at some intermediate value between (and including) the two extremes.
\\
C \& H discuss that with their normalisation metastable mBHs
with lifetimes of many years exist, but only for very large
values of L approaching the experimentally excluded range L$>$10$^{-4}$ m\cite{kapner}.
It can be easily shown that
with normalisation eq.(\ref{norm2}) 
mBHs are quasistable for
all possible values of L\footnote{i.e. the range from
$\hbar$/(c M$_5$) to 10$^{-4}$ m}.
\subsection{Summary}
Summarising, mBHs can be ``quasistable'' in {\it scenario 3}
(in the sense of lifetimes exceeding O(msecs)), without introducing
highly implausible ``ad hoc'' assumptions. 
Because G \& M concluded that 5-dimensional and
sufficiently stable mBHs might accrete matter
at an extremely fast rate (growth rate much below a second, see below section \ref{hdes})
quasistable mBHs are potentially dangerous. 
In contradistinction to G \& M's {\it scenario 2},
my {\it scenario 3} is plausible as a fundamental theory of
mBHs and therefore 
an astrophysical (or other empirical)
exclusion of the existence of such mBHs 
is a ``critical safety guarantee'' rather than an 
``additional check of their benign nature'' as
it was characterised by G \& M for {\it scenario 2}.

\section{A potential threat from microscopic black holes 
Hawking-radiating at the Eddington limit within {\it scenario 3}}
\label{hdes}
\subsection{Introduction}
The mBHs in {\it scenario 3} emit Hawking radiation, and according
to eq.(\ref{lumi5}) the emitted power rises linearly with the their mass.
Might this radiation be more dangerous than the mechanical action of the accretion?
Unfortunately it turns out that this might be the case for certain parameter choices.
In this section I choose possible and not fine-tuned parameters to study and illustrate
the nature of the possible risk.
\subsection{The nature of ``Hawking radiation risk'' for one exemplary choice of input parameters}
\label{hdes2}
For purely illustrative purposes - as one
concrete instantiation of {\it scenario 3} - I set L=10$^{-7}$ m below. Let us further assume that
M$_N$ = 1.9 $\times$ 10$^5$ kg, a value intermediate between
the one given by first and second normalisation (section \ref{ch}).
According to eq.(\ref{lumi5}) mBHs would then have a lifetime of about 
2 seconds. A collider-produced mBH that has been captured and slowed
down to thermal velocities, accretes and quickly grows by the ``subatomic
accretion mechanism'' (the sucking in of particle within an
atom by the mBH) characterised in section 4.2 of G \& M.
According to G \& M's eq.(4.22) it will take about 0.15 msec 
until the so called ``electromagnetic radius'' reaches atomic sizes\footnote{
A conservative thermal velocity of 1500 m/sec was used to convert the units in 
eq.(4.22) to a time.}.
Thereafter the accretion is well described as Bondi accretion 
(the sucking in of whole atoms by the mBH) and
according to eq.(4.40) in G\& M it will take about 2.2 msec until 
the mBH's Schwarzschild radius reaches L=10${-7}$ m at a mass of 0.54 kg. 
The further evolution 
of the mBH's shape and size in the ``transitional region'' between
5 and 4-dimensional behaviour (see section \ref{ch}) is not well
understood. For simplicity 
I will assume that the radius remains constant at L (a radius increase
logarithmic with the mBH's mass\cite{gid03} would not change the results
appreciably.). For the input parameters chosen in this subsection, 
eq.(4.31) of G \& M predicts
an increase of the mBHs mass at a rate of 1.9 $\times$ 10$^4$ kg/sec.
It will then take about 20 $\mu$sec until its mass reaches about 1 kg.
At this mass the luminosity of the mBH is predicted by eq.(\ref{lumi5})
to be 5.1 $\times$ 10$^{16}$ W or a mass equivalent of dm/dt = 0.57 kg/sec.
It is easy to verify that the five-dimensional Eddington limit (eq.(B.25) of G \& M) 
\begin{equation}
dM/dt = {{2.44 \times 8 \pi m_p R_B c_s^2}\over{\eta \sigma c}} 
\label{eddington}
\end{equation}
has the same magnitude for an efficiency $\eta$=1.
Here m$_p$ is the mass of the proton, R$_B$ the Bondi radius (4.1 mm for our
parameters), c$_s$ the
velocity of sound in the interior of Earth (5200 m/sec) and $\sigma$ the
Thomson cross section.
\\
Therefore the radiation pressure of this Hawking radiation is intense enough to 
limit the mass of accreted matter to the mass-energy radiated away: dm/dt = dM/dt i.e. the mBHs accretes
at the 5-dimensional Eddington limit. All accreted mass
is then reradiated, and the mBH's mass remains constant on average. G \& M discussed the possibility 
of a radiation-limited accretion and
excluded it, but only because in their {\it scenario 2} the Hawking radiation is completely switched off.
\\
For the next 3 $\times$ 10$^{17}$ years, a time span vastly exceeding 
the life time of our sun as a normal star, the mBH will radiate at the
quoted, constant luminosity. 
The power of 5.2 $\times$ 10$^{16}$ W is 1300 times larger 
than the total geothermal power emitted by Earth\cite{geoth}, and only 3 times less
than the total power Earth receives from the sun. The radiated power exceeds the total
seismic power if the Earth by an estimated factor of many millions\cite{seis}. 
17000 metric tons of ambient matter would be converted to radiation each year.
While the exact
phenomenology provoked by such a mBH
accreting at the Eddington limit remains to be worked out, eventually catastrophic
consequences due to global heating on an unprecedented scale and 
global-scale earth-quakes would seem certain.
\subsection{Can the risk be ruled out with astrophysical arguments?}
Disturbingly the effects of such a mBH on a white dwarf or neutron 
star would be negligible. Assuming the same mBH parameters
as above and the theory of section 7 in G \& M, the luminosity 
of the mBH accreting at the centre of a white dwarf is predicted 
to be 5.9 $\times$ 10$^{19}$ W
or a fraction of 1.5 $\times$ 10$^{-7}$ of the solar luminosity. 
This is about 10$^4$ times smaller than the cooling rate of 
white dwarfs in G \& M's sample\cite{gm,isern} and
thus cannot be detected\footnote{G \& M find that many mBHs are produced
in white dwarfs in the course of time. However, these mBHs will also 
tend to merge over time, so that the total number of black holes in a
given white dwarf might remain small. This question needs further study.}. 
The accretion time of a white dwarf would exceed their present age by a large
factor of $>$ 10$^{10}$.
Therefore no conclusions about mBHs can be drawn from the observed existence 
of such objects with ages exceeding a billion years.
The conditions for a neutron star would be similarly unspectacular.
Therefore the astrophysical argument
of G \& M fails to exclude the existence of mBHs in {\it scenario 3} that are
dangerous not because they accrete the whole Earth but 
because of their intense Hawking radiation.
\subsection{A local accident at CERN?}
\label{cern}
The luminosity of a mBH accreting at the Eddington limit with the
parameters assumed above corresponds to 12 Mt TNT equivalent/sec\cite{seis}, or
the energy released in a major thermonuclear explosion per second.
If such a mBH would accrete near the surface of Earth the damage they
create would be much larger than deep in its interior.
With the very small accretion timescale ($\ll$ 1 second) that was found with the
parameters in subsection \ref{hdes2},
a mBH created with very small (thermal or subthermal) velocities in a collider
would appear like a major nuclear explosion 
in the immediate  vicinity of the collider. 
The risk from collider-produced black holes is not necessarily an 
Armageddon, but could be a locally contained catastrophe.
\subsection{Conclusion}
If black holes are described by {\it scenario 3} the Hawking radiation
from collider-produced black holes might be dangerous.
The input parameters used in this section were only one example.
It can however be shown that there is a wide range of values for L and M$_N$ that lead to dangerous mBHs
accreting at the Eddington limit with various luminosities.
\\
In general the example developed above demonstrates that 
widely held intuitions - namely that
accretion by mBHs must be extremely slow\cite{bleicher}, and that events which are catastrophic for
Earth must also be for compact stellar objects (necessary for any safety argument based on such
objects)
- are insufficient as safety
guarantees. 

\section{Does the observed existence of old white dwarfs with a low magnetic
field rule out ``dangerous'' stable black holes? - A gap in G \& M's
exclusion of their {\it scenario 2}}
\label{danger}

In this section I point out a fundamental weakness of G \& M's argument
that cosmic rays impinging on white dwarfs rule out the existence of
dangerous mBHs. This argument puts into question whether {\it scenario 2}
as defined in the introduction is really ruled out by existing
astrophysical observations.
\\ 
In the text following their
eq. (E.2) G \& M formulate the following assumption: 
\begin{equation}
M_{\rm min} > 3 \ M_5
\label{constr} 
\end{equation}
Thereby G \& M introduce the
assumption that mBHs in general have a minimal mass M$_{\rm min}$ that exceeds
the new Planck scale by at least a factor of 3.
This constraint is motivated by the fact that the
thermodynamical, semiclassical treatment of mBHs in their {\it scenario 2}
is expected to be reliable within this mass range.
This is certainly a most reasonable argument for all  
purposes of pure research,
e.g. when predicting collider signatures etc..
However, it does not mean that mBHs below 
M$_{\rm min}$ cannot be produced. It rather means that we are presently
unable to reliably predict the behaviour of such mBHs\footnote{
In a previous paper\cite{gidthom} Giddings wrote: ``For masses of
order the fundamental Planck scale [i.e. M$_5$] there is no control
over quantum gravity effects which are likely to invalidate
the semiclassical ... picture.''}.
\\
This fact raises a fundamental doubt about G \& M's exclusion 
of ``dangerous mBHs'' by way of observationally constraining the age
of a certain class of white
dwarfs. This exclusion depends on their careful and
detailed demonstration in their section 5 that ``dangerous''
mBHs are stopped in white dwarfs after their production in collisions
of cosmic rays. However, this demonstration is based on an assumed validity of
the semiclassical
approximation. mBHs deep in the ``quantum gravity'' regime (violating
eq.(\ref{constr})) might have smaller scattering cross section
than expected in the semiclassically and escape white dwarfs, just
as they could escape ordinary stars
\footnote{This is a real concern because the
``safety critical'' black holes in theories with 2 extra dimensions 
cannot be excluded anymore if their scattering cross section 
is smaller by less than a mere factor 10 than semiclassically expected (see fig.2 of G \& M\cite{gm}).}.
This would void G \& M's exclusion of the existence of potentially ``dangerous''
black holes. 
\\
Concluding, G \& M did not demonstrate with reasonable certainty that white dwarfs 
stop cosmic-ray produced mBHs in general. Their exclusion of dangerous
mBHs thus remains not definite.

\section{Conclusion}
\label{concl}
\subsection{Summary of reasons for concern}
I showed in section \ref{hdes} that within {\it scenario 3}
(as outlined in section \ref{scen3}),
mBHs produced at a collider can be
captured by Earth and accrete at the Eddington limit. Thereby they might emit
Hawking radiation that might be dangerous to 
Earth as a whole or the inhabitants of CERN
and its surroundings. The astrophysical argument by G \& M\cite{gm} and Koch et al.\cite{koch}
does not exclude this scenario, because it allows for
lifetimes of white dwarfs and neutron stars
with Eddington accreting mBHs (section \ref{hdes}) far exceeding the
age of the universe. Such stars could still exist, even
if they harboured a microscopic accreting black hole since the dawn of time.
\\
Moreover - in section \ref{danger} - I outlined another independent gap
in the ``astrophysical exclusion'' of potentially dangerous
mBHs.
\\
Thus, at the present stage of knowledge there is a definite residual risk from mBH production
at colliders.
This final conclusion differs from the one drawn by G \& M.
This is not because of any disagreement over any specific conclusion
of their excellent paper. 
Rather the difference is the sole result of either 
employing the alternative, physically plausible {\it scenario 3}
for the physics of mBHs or
including parameter regions in which mBHs 
are not expected to be well described by a semiclassical
approximation of quantum gravity into the safety analysis.
\subsection{Proposal for risk mitigating measures}
It is not the aim of the present paper to recommend or discuss 
consequences for the future operation of colliders comprehensively.
Here I just put up for further discussion
three feasible measures for risk mitigation,
{\it at least in the start up phase of LHC}:
\\
1. Increase of collision energy by reasonably small factors (say, 2)
in one step and proceed to higher
energies only after excluding any indication for potentially dangerous
processes.
Currently it is planned to perform the
first production runs at LHC at an energy more than five times higher
than previously reached\cite{comm}\footnote{Due
to technical problems, very recently a 
reduction of this value to 3.5 times 
higher has been decided\cite{cho}.}. This might result in the copious
production of completely novel states, which production
was exponentially suppressed at the previous energies. 
``Proceeding in small steps" mitigates this risk.
\\
2. No operation in which no or only a very tiny fraction
of events are analysed. Currently it is planned to
eventually record and analyse only a fraction of 10$^{-7}$ of
all events\cite{stapnes}. This is the equivalent of entering new
territory and to be on the lookout only for the
interesting but not the potentially dangerous.
\\
3. Safety considerations influence the trigger and
operational procedures. Meta stable black holes  
might not yield very spectacular events, but it seems desirable
to ensure that their presence is immediately and reliably detected. 
An immediate interruption of operation and detailed
off-line study of the event might be a possible
risk-mitigating measure.
\\
Measures 1. only reduces the risk if measure 3. is also taken.
\\
To take such safety measures would not
exclude but reduce any remaining risk. Methodologically
similar measures have been taken in other areas of fundamental
research under analogous
circumstances, e.g. in biotechnology\cite{asilomar_sum}.

\section{Acknowledgements}
I thank G.t'Hooft, M.Jarnot, M.Leggett and S.Pezzoni for
critical comments on a previous version of the present manuscript.
I thank R.Casadio for his patient and helpful explanations of his theory.

\section{Appendix - Answer to the manuscript ``Comments on claimed risk from metastable black holes''
by S.Giddings \& M.Mangano}
\label{gm_crit}
In a preprint\cite{gm_resp} S.Giddings and M.Mangano (G \& M) raise 
five objections against the conclusions of the present manuscript. 
I answer them below and raise, as a sixth point, an omission in their report.
\\
1. 
They ``find a negligible power output 
[from a black hole with the properties described in 
section \ref{hdes} of the present manuscript]
of the size 0.1 $\mu$W, differing
by a factor of 10$^{23}$ from [my] claim.''
\\
My reply:
\\
G \& M employ their
eq.(1)\cite{gm} - i.e. they assume  {\it scenario 1} (see section \ref{risk_scen} of 
my paper) - to calculate the power output of a 5-dimensional 
microscopic black hole with a radius of
10$^{-7}$ m within the canonical thermodynamic description 
of microscopic black holes. They correctly find it to be
23 orders of magnitude smaller than the power output calculated in section
\ref{hdes} of my paper for a black hole of the same size and conclude
that my result is erroneous. They claim that I mistakenly applied their eq.(1) 
``written in terms of the mass using the four-dimensional relationship
between radius and mass''.
\\
However it is the crucial point of my paper
to introduce {\it scenario 3} that does {\bf not} 
employ their eq.(1) or my eq.(\ref{lumi4})
i.e. a thermodynamic, canonical
description to calculate the power output of the black hole.
Rather, following Casadio \& Harms\cite{casadio},
it exclusively employs my eq.(\ref{lumi5}) 
(eq. (28) in \cite{casadio})
which has a qualitatively different form.
Following Casadio \& Harms \cite{casadio} (sentence after
their eq.(28)),
my eq.(\ref{lumi4}) was only used to normalize 
the luminosity eq.(\ref{lumi5})
at the mass M$_N$. 
\\
This objection criticises something I did not do and did not intend to do.
Therefore it does not
apply to my paper.
Implicitly G \& M simply insist, here and in the
third objection below, on a canonical treatment of mBHs, i.e.
to employ their {\it scenario 1}, without stating any reason 
for implicitely ruling out {\it scenario 3}.
\\
2. They claim that ``even ignoring the inconsistency'' above,
the bounds established in Giddings \& Mangano\cite{gm} on 
Eddington limited accretion (like established in their eq.(B13))
still apply. 
\\
I reply:
\\
As already pointed out in the 4th sentence after my
eq.(\ref{eddington}) all results on Eddington limited
accretion of G\&M were 
derived for their {\it scenario 2} with ``switched off''
Hawking radiation and therefore
do not apply for the case ``Hawking-radiation limited'' accretion at
the Eddington limit, considered in my section \ref{hdes}.
\\
3. They feel that a ``serious difference between the
microcanonical picture and the usual Hawking calculation 
appears implausible in the large black hole region''
I considered (following Casadio \& Harms).
\\
I reply:
\\
G \& M give reason for this evaluation - which, I note, is also in flat
contradiction to the work of Casadio \& Harms\cite{casadio} - so I 
need to wait for their promised ``further comments''
to take a position. 
\\
In light of our poor understanding of black-hole evaporation
in general (see quote below) I feel that in any case it will be difficult
to rule out such a serious difference with reasonable certainty.
\\
4. G \& M rightly point out that in my quote:
``...at each point where we encountered an
uncertainty, we have replaced it by a conservative {\bf or} ``worst case'' assumption''.
The bold ``or'' was missing. I corrected this oversight in the 
revised version.
\\
5. G \& M further propose that I quote from an abstract of a talk by W.Unruh\cite{unruht},
in addition to the references of my paper. I hereby accept 
their suggestion. An excerpt from Unruh's abstract
reads: ``{\it ...Black Hole evaporation is one of the most puzzling
features of gravity and quantum theory. The derivation by Hawking is nonsense, 
in that it uses features of the theory in regimes where we know the theory 
is wrong. Analog models of gravity have given us a clue that despite the 
shaky derivation, the effect is almost certainly right. 
Where then are the particles in black hole evaporation really created?...''} 
\\
From this quote I conclude: theories
with extra dimensions robustly predict the existence
of microscopic collider-producible black holes
and Hawking radiation. But the detailed decay properties 
presently remain very uncertain. It then seems important to study
the safety issue assuming
plausible, literature-based alternatives to the standard thermodynamical treatment 
of Hawking radiation. This is the aim of my paper. 
\\
6. Finally G \& M's comment did not address section \ref{danger} of
the present manuscript in which I argue that their exclusion
of dangerous mBHs is not completely 
definite for a general, simple
reason, completely independent of
the above arguments. 
\\
I stand by my general conclusion that there is a residual
catastrophic risk from metastable microscopic black holes
produced at particle colliders. 

\section{Appendix - Comment on the manuscript ``On the possibility
of Catastrophic Black Hole Growth in the Warped Brane World Scenario
at the LHC'' by Casadio et al.}
\label{cas_paper}
\subsection{Identity of the general approach of Casadio et al. and the present paper}
In a recent important preprint\cite{cfh}, Casadio et al. studied
the same question as the present
manuscript: might collider-produced black holes grow catastrophically?
Based on their own earlier groundbreaking\footnote{Casadio \& Harms (2002) Ref.\cite{casadio} 
is in the ``top cite 100+''
category of High-Energy Physics Literature Database ``Spires'' 
and one of the earliest and most frequently cited
papers on microscopic black holes.} work\cite{casadio} 
they assume the same theoretical approaches for the behaviour
of quantum black holes as I did.
For the calculation of accretion (the sucking in of matter by gravitational
forces)
we both used the results of Giddings \& Mangano\cite{gm}. 
As a consequence, e.g., the
calculated lifetimes of black holes in the left-handed panel of
their fig. 4\footnote{My discussion refers exclusively to their final version 2 of their manuscript.
}
are reproduced by my eq.(1).
We will see below that {\it with identical parameter choices} their 
results on black-hole accretion also broadly agree with the
ones obtained by the methods employed in the present manuscript\footnote{
Casadio et al. modelled the evolution of quantum black holes numerically,
whereas I merely used analytical approximations from Giddings \& Mangano\cite{gm}
for the accretion.}.
\\
Casadio et al.\cite {cfh} and I 
study the evolution of collider-produced 
metastable black holes under the same general assumptions
that correspond to {\it scenario 3} discussed
in the introduction to the present manuscript. This disagrees
with the criticism of the CERN team\cite{gm_resp} of my manuscript that
basically denied the possibility of my {\it third scenario} (see above section \ref{gm_crit}). 
Still, Casadio et al.
``argue against  the possibility of catastrophic black hole growth'' whereas I find a ``residual risk'' for it. Let us
examine in detail how their parameter choices differ from mine. 
 
\subsection{Analysis of parameter choices of Casadio et al.}
\label{casa2}
\subsubsection{The crucial difference is that Casadio et al. assumed that M$_c$ $<$ 10$^4$ kg}
For the parameter values generally
assumed by Casadio et al.\cite{cfh}\footnote{Given in their section II.C and IV.B.} (new Planck scale M$_5$ = 1 TeV/c$^2$,
atomic number accreted nuclei A=53,
material constant K=224 J/m$^2$, number of dimensions D=5) 
I find from eq.(4.22) of Giddings \& Mangano\cite{gm} 
that a black hole needs a time t$_{\rm accr}$ $\approx$ 0.6 msec to reach a
size\footnote{more precisely: to reach an electromagnetic capture
radius ``R$_{\rm EM}$''.} of 1 $\buildrel _\circ \over {\mathrm{A}}$ 
via ``subatomic accretion'' (i.e. via
gravitationally sucking in elementary particles 
within one atom)\footnote{A thermal velocity of 1500 m/sec
of the ambient atoms was assumed.}.
When the expected lifetime of the black holes ``t$_{\rm decay}$''
exceeds t$_{\rm accr}$ the black hole can reach this size before
decaying
and at later times
``Bondi accretion'' (the sucking in of whole atoms) takes
over which is even faster than subatomic accretion for the
assumed parameters (its time scale is about 0.1 msec)
and leads to exponential, catastrophic growth.
However, one sees from fig. 4 in Casadio et al.\cite{cfh}
that t$_{\rm decay}$ remains below t$_{\rm accr}$ 
just for the range of critical masses M$_c$ $<$ 10$^4$ kg
considered by Casadio et al.\cite{cfh}
(M$_c$ is a ``free
parameter'' of the theory of Casadio \& Harms\cite{casadio},
i.e. its value is not fixed). Therefore, I completely agree with
the technical conclusion of Casadio et al.\cite{cfh}: {\bf If}
M$_c$ $<$ 10$^4$ kg, collider-produced black holes
never reach the Bondi regime and growth to catastrophic
sizes is not expected.
\\
However, the crucial question clearly is: can M$_c$ exceed
a critical value of about 1.3 $\times$ 10$^4$ kg, which would make 
the predicted liftime t$_{\rm decay}$
of quantum black holes in the theory of Casadio \& Harms
exceed the accretion time t$_{\rm accr}$, needed to reach the catastrophic
Bondi regime? In section \ref{hdes} of the present paper
I chose M$_c$ = 1.9 $\times$ 10$^5$ kg (called
M$_N$ there), thus implying
that it can.
On the other hand
Casadio et al.\cite{casadio} 
call values of M$_c$ exceeding 
10$^4$ kg ``physically unreasonable''. Let us analyse how
they reached this conclusion, that directly led them
to the exclusion of a catastrophic risk.
\subsubsection{Can M$_c$ $>$ 10$^4$ kg be excluded?}
Casadio et al.\cite{cfh} employed their eq.(18) (in section II.C) 
and eq.(25) (in section III.A)\footnote{The derivation of eq.(25) 
is based on the requirement that the so called ``electromagnetic
capture radius'' of a black hole
with a mass of M$_c$ shall always be smaller than L. I do not
understand what plausible connection there could be between
the capture radius and black hole properties. However, in view
of our limited understanding of black holes 
with radii near L their eq.(25)
might accepted; besides
their eq.(16) and eq.(18), as another ``possible choice'' for the
estimation of M$_c$.} as the ``defining condition'' to estimate
the value M$_c$ as a function of the warping scale ``L'', a free parameter
in the RS2 theory of extra dimensions\cite{randall}\footnote{
L specifies the length over which a fourth spatial dimension is
significantly curved.}. Using eq.(18) (eq.(25)) I find that
M$_c$ exceeds the critical value of M$_c$ of about 1.3 $\times$ 10$^4$ kg
where the growth becomes catastrophic
if L $>$ 15.3 $\mu$m (L $>$ 12.2 micrometer).
As pointed out by Casadio et al. the current best experimental
current limit on L is L $<$ 44 $\mu$m\cite{kapner}. This limit is 
a 95 \% upper limit, i.e. there is a chance of 5 \% that
L actually exceeds 44 $\mu$m. I conclude therefore that there
is certainly no comprehensible reason to exclude values of L
between 15.3 and 44 $\mu$m. 
\\
In section II.C 
directly after pointing out that $L\lesssim 44\,\mu$m,
Casadio et al. continue: ``For example setting L $\simeq$ 1 $\mu$m
... gives ... M$_c$ $\simeq$ 10$^2$ kg.'' This is correct,
but in the rest of their paper, the latter value is suddenly and
without giving any reason, treated as an {\it upper limit}
on M$_c$. 
To me only the value of M$_c$ obtained with
44 $\mu$m is comprehensible as the upper limit.
Inserting L=44 $\mu$m into eq.(18) (eq.(25)) of Casadio et al.
one obtains M$_c$=1.1 $\times$ 10$^5$ kg (2.2 $\times$ 10$^6$ kg) 
as the presently valid upper limit. 
The predicted
lifetime of the black hole assuming M$_c$ = 1.1 $\times$ 10$^5$ kg
t$_{\rm decay}$= 0.3 sec, exceeds the
accretion time scale t$_{\rm accr}$=0.6 msec by a factor of 50: in this case growth is
catastrophic. 
\\
Moreover - as pointed out by Casadio et al.\cite{cfh} - within the
uncertainties of their theory, eq.(16) is, quote, ``another possible choice''
for the estimation of M$_c$. 
Inserting L=44 $\mu$m into eq.(16) of Casadio et al.
one obtains the huge value M$_c$=5.9 $\times$ 10$^{22}$ kg. 
With this choice for M$_c$ growth
is found to be catastrophic for all reasonable
values of L\footnote{I discuss these alternative choices in 
my section \ref{ch}. I call Casadio et al.'s M$_c$ ``M$_N$'', 
their eqs.(16),(18) correspond to my eqs.(\ref{norm2}),(\ref{norm1}).
My choice of M$_c$=1.9 $\times$ 10$^5$ kg in section \ref{hdes2}
is justified when assuming eq.(16) in Casadio et al..}.
\\
Until reasons are brought forward to exclude 
values of L exceeding 15.3 $\mu$m and the use
of their eq.(16) to estimate M$_c$, catastrophic
growth remains a possibility.

\subsection{The ``astrophysical'' safety argument of G \& M in the light of
the model of Casadio et al.}

Giddings \& Mangano\cite{gm} did not categorically exclude the possibility
of catastrophic growth, but argued that if it happened certain
astrophysical objects would be {\it completely} consumed by 
cosmic-ray produced black holes on time scales
smaller than their age, which is contrary to observations.
In the current manuscript (section \ref{hdes}) I discussed a mechanism 
that invalidates this argument in the case of metastable
black holes (``{\it third scenario}'') because their growth 
by accretion is eventually stopped at intermediate
sizes via their emission of Hawking radiation. 
The question of which other mechanisms
in the evolution of quantum black holes in a microcanonical
framework might avoid G \& M's basic safety argument
needs further scrutiny.

\subsection{Summary}

Summarizing, the work of Casadio et al.\cite{cfh} confirms the
validity and plausibility of my ``{\it scenario 3}'' based
on their microcanonical description of metastable black holes. This scenario
was completely excluded from the safety analysis conducted by CERN\cite{gm}
\footnote{Even in a revised version of their report 
with ``updated references''(\cite{gm} v2 of the arXiv version) that
was prepared after its authors had quoted Ref.\cite{casadio} in 
Ref. \cite{gm_resp}) 
they fail to even quote Casadio \& Harms\cite{casadio}
or indeed any paper on the microcanonical approach to
black hole physics. The revised manuscript 
still contains sentences such as ``Thus, on very general grounds such [with a mass
of about 10 TeV] black
holes are expected to be extremely short-lived... t$_{\rm D}$ $\approx$ 10$^{-27}$ sec...''
which completely fly into the face of Casadio et al.'s conclusions: they argue that
lifetimes exceeding t$_{\rm D}$ by about 17 orders of magnitude are possible.}
without providing any reason for this, at least up to now (see third bullet point
in section \ref{gm_crit}).
Casadio et al. correctly and usefully identify the range
of a crucial theoretical parameter M$_c$
for which black-hole growth is not catastrophic,
but offer no argument of how to exclude that the real
M$_c$ lies outside this range.
\\
Acknowledgement
\\
I thank M.Jarnot, M.Leggett, L.Lueptow, E.Penrose and S.Pezzoni for helpful remarks
on previous versions of this section and R.Casadio and B.Harms for 
answering some questions about their manuscript.

\end{document}